# Machine Learning Based Network Vulnerability Analysis of Industrial Internet of Things

Maede Zolanvari, *Student Member, IEEE,* Marcio A. Teixeira, *Senior Member, IEEE*, Lav Gupta, *Senior Member, IEEE,* Khaled M. Khan, *Member, IEEE,* Raj Jain, *Life Fellow, IEEE*

***Abstract*— It is critical to secure the Industrial Internet of Things (IIoT) devices because of potentially devastating consequences in case of an attack. Machine learning and big data analytics are the two powerful leverages for analyzing and securing the Internet of Things (IoT) technology. By extension, these techniques can help improve the security of the IIoT systems as well. In this paper, we first present common IIoT protocols and their associated vulnerabilities. Then, we run a cyber-vulnerability assessment and discuss the utilization of machine learning in countering these susceptibilities. Following that, a literature review of the available intrusion detection solutions using machine learning models is presented. Finally, we discuss our case study, which includes details of a real-world testbed that we have built to conduct cyber-attacks and to design an intrusion detection system (IDS). We deploy backdoor, command injection, and Structured Query Language (SQL) injection attacks against the system and demonstrate how a machine learning based anomaly detection system can perform well in detecting these attacks. We have evaluated the performance through representative metrics to have a fair point of view on the effectiveness of the methods.**

## I. INTRODUCTION

The primary concept of the Industrial Internet of Things (IIoT) is to take advantage of IoT technology in the industrial control system (ICSs). ICSs are an integral part of critical infrastructures and have been utilized for a long time to supervise industrial machines and processes. They perform real-time monitoring and interacting with the devices, real-time collection and analysis of the data, and also logging of all the events that happen in the industrial systems. Utilizing IoT technology in these systems enhances the network intelligence and security in the optimization and automation of industrial processes.

The supervisory control and data acquisition (SCADA) system is the largest subset of an ICS. It provides a graphical user interface (GUI) through its human-machine interface (HMI). The HMI makes it easy for the operators to observe the status of the system, interact with the IIoT devices, and receive alarms indicating abnormal behaviors. A general scheme of SCADA systems is shown in Figure 1. As shown in this figure, these systems consist of four different sub-systems; I/O network, supervisory control, control network, and corporate network. I/O network consists of the deployed IIoT devices (including sensors and actuators) in the industrial process. Supervisory control is the main sub-system responsible for securing, controlling, and monitoring the IIoT devices. The control network includes programmable logic controllers (PLCs) that directly sense and manage the physical processes. Since the sensors and actuators cannot communicate directly, PLCs are used to collect the sensed data and send commands to the actuators. Finally, the corporate network consists of servers, computers, and other users connected to the network for other general services such as file transfer, website hosting, mail servers, resource planning, etc.

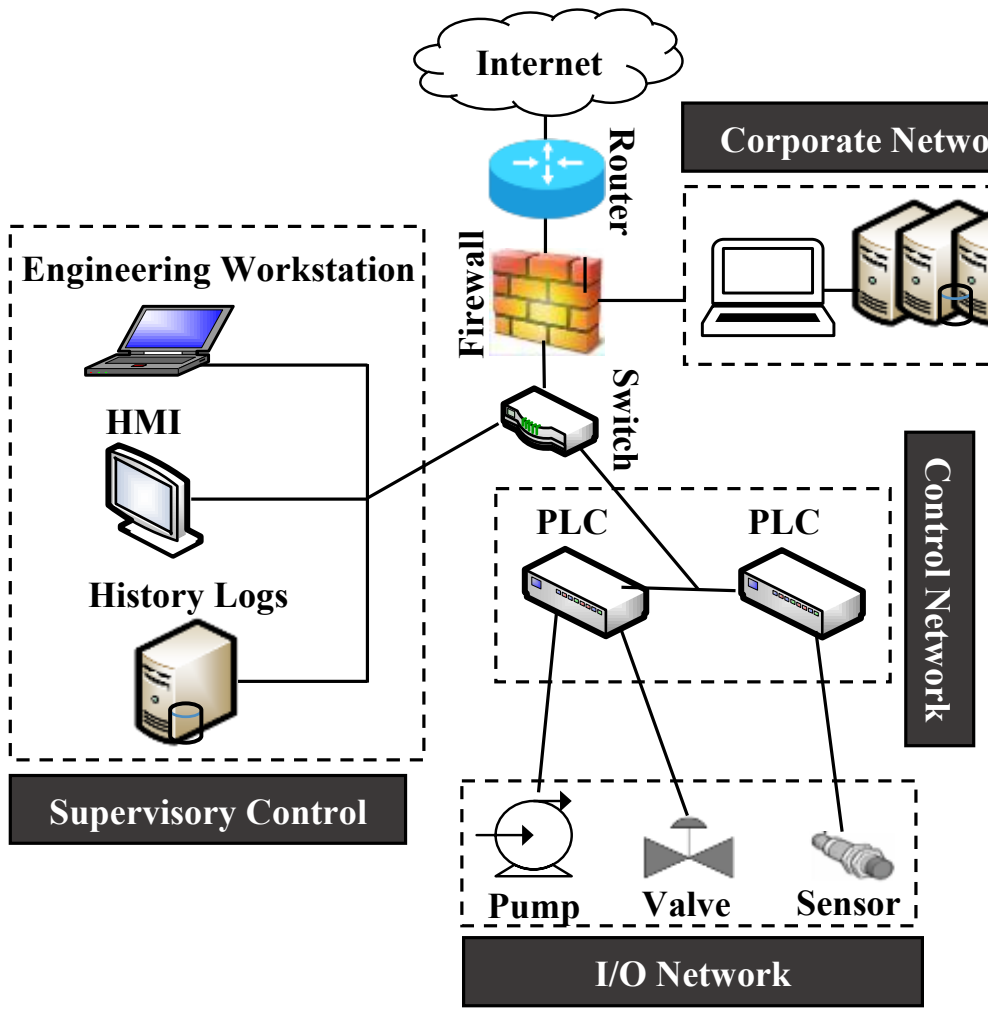

Figure 1. SCADA Architecture

ICSs are mostly mission-critical systems with high-availability requirements. Their continuous operations lead to producing a huge amount of data that can be managed through big data analytics. In the past, these systems were standalone and isolated from the world, making them unsusceptible to external malicious attacks. Recently, increased connectivity of ICS with corporate networks and utilization of Internet communications to transmit the information more conveniently have rendered these systems vulnerable to malicious attacks. Due to the sensitive nature of many industrial applications, security has become the primary concern in SCADA systems. More specifically, lack of security considerations in their communication protocols directly compromises the availability, safety, and reliability of these systems. Our work in this area shows that machine learning based solutions can introduce new countermeasures to secure these systems.

The main contributions of this paper are as follows:

- Describing the four most popular IIoT protocols, along with their main communication network vulnerabilities (Section II).
- Conducting a vulnerability assessment of IIoT systems, identifying common malicious threats, analyzing the severity of the associated risks, and also studying the applicability of machine learning techniques to counter these threats (Section III). This is an essential step to prioritize required mitigations.
- Reviewing the research papers that have designed efficient machine learning based intrusion detection systems (IDSs) for SCADA systems (Section IV).
- Presenting our own case study, elaborating on the testbed designed at Washington University in St. Louis, to perform real-world SCADA operations, carrying out attacks (that have not been implemented either for the under-study or any other IIoT system by other researchers) against the system, and applying machine learning based IDS to tackle the intrusion detection problem, studying the most important features in identifying the attack traffic from normal (Section V).

## II. IIOT COMMUNICATION PROTOCOLS

There are several IIoT data transmission protocols used in the SCADA systems. However, most of these protocols have been designed without accounting for cyber risks or security mechanisms to counter them. The legacy of SCADA started with Modbus communication protocol which is still the most widely employed protocol in these systems. Recently, there has been a trend in moving towards newer protocols such as Building Automation and Control Network (BACnet), Distributed Network Protocol version 3 (DNP3) and Message Queuing Telemetry Transport (MQTT). All four are open protocols. Modbus was developed as a SCADA-vendor specific protocol in 1979 [1]. BACnet and DNP3 are standard protocols that were published in 1995 and 1993, respectively [2]. MQTT was developed in 1999 [3].

In this section, we study these four popular SCADA communication protocols along with their main security vulnerabilities. Since the main focus of this paper is on the network susceptibilities of the IIoT, this section has been included to show where each communication protocol is most vulnerable.

### A. *Modbus*

Modbus is one of the earliest and the most commonly used protocol in the SCADA systems. The communication is serial and based on master-slave configuration. Master (e.g., HMI) is the device requesting the information and slave (e.g., PLC) is the one supplying the information. The master can also write data on the slave’s memory registers. In a standard Modbus communication network, there is one master that can have up to 247 slaves, each with a unique identification code (ID). There are four tables stored in the slave device, two for storing digital data and two for numerical analog data.

Modbus does not provide confidentiality, authentication or integrity. Because all Modbus traffic is communicated in clear text (no encryption is provided), it lacks confidentiality, and the content of the packets can be easily seen using a sniffer tool. Modbus does not provide any public/private key management, which leads to lack of authentication as well. Also, there is no sufficient security check mechanism on the traffic, which makes it easy for the attacker to compromise the integrity of the data. Moreover, a flooding attack can interrupt the operation of the system and limit its availability.

### B. *BACnet*

BACnet was primarily designed for building automation and control systems. This standard is currently under ASHRAE Standing Standard Project Committee (SSPC) 135. BACnet, like most of the other industrial communication protocols, was not designed with security considerations.

This standard provides several communication options, such as Ethernet, token-passing, master-slave, or point-to-point connections [4]. In SCADA, it is more common to use the master-slave mode of BACnet.

Due to lack of proper mechanisms for data confidentiality, reconnaissance attack is feasible. This protocol does not provide authentication procedures either. A few cryptography mechanisms, e.g., DES (Data Encryption Standard) and AES (Advanced Encryption Standard), have been included in the new versions of the BACnet standard. However, they are almost never utilized in industrial systems to maintain compatibility with the existing devices. Even in new green-field installations and in general, implementing encryption adds communication delays and large overheads in the system. The scan cycle times of PLC and HMI are usually on the order of milliseconds, and it is not yet feasible to accommodate data encryption in the system since encryption and decryption require a longer time process. In addition, denial of service (DoS) attacks can be conducted to target the system availability and halt the service [5].

### C. *DNP3*

DNP3 is another standard network protocol used in SCADA systems. It was originally designed to be very reliable, but it does not provide sufficient security mechanisms. As a result, most of the DNP3 devices lack authentication, encryption, and access control. DNP3 covers the four OSI layers: network layers, application layer, data link layer, and physical layer. The communication can happen in point-to-point mode but mostly happens in master-slave configurations, and it can include multiple slaves and multiple masters.

No message authentication is deployed in this protocol, and hence data integrity is at risk. Eavesdropping and spoofing are easy as there is no encryption and the data is sent in clear text. DoS attacks can also easily impact the system’s operation [6]. The latest version of this protocol that was published by IEEE in 2012, provides secure authentication, using IEC 60870-5 standard [7]. Through this standard, which has been developed for control systems, the authentication is provided using digital signatures. However, utilizing public key infrastructure (PKI) in IIoT devices is not feasible yet. The complexity that PKI adds cannot be handled by the simple IoT devices. Further, exchanging the keys, updating the keys, issuing or revoking certificates and other complexities that come with using PKI will add a huge delay in the system's performance.

### *D. MQTT*

MQTT is an open standard under OASIS and is based on publish/subscribe configuration. This network protocol has been very popular in the IoT domain because MQTT messages are very simple and lightweight. Recently, MQTT has been increasingly adopted in the ICSs due to its suitability for remote sensing and control.

MQTT's topology consists of clients and brokers. At any particular time, each client can be either a publisher or a subscriber based on whether they are requesting or supplying data. A broker is an intermediate device between the publishers and the subscribers to filter out the published data and send them to their subscribers. Each broker can handle thousands of clients, which helps the system with scalability.

No encryption method has been implemented in MQTT, but TLS/SSL (Transport Layer Security/ Secure Sockets Layer) can be applied on the underlying TCP/IP (Transmission Control Protocol/ Internet Protocol) to provide an encrypted pipeline for the MQTT messages. However, since this requires a high level of complexity on the clients, it is not practical to use in the IIoT devices. Another main security drawback of the MQTT that originates from its topology is that, if an intruder steals the identity of a client, it will have access to all other clients' data, and not only that specific victimized client. On the other hand, the broker can be designed to ask each client for their username and password to allow them to join the network. However, these credentials will be transmitted in clear text, if no form of encryption is utilized. For data integrity, MQTT can provide MAC (message authentication code) techniques such as HMAC (hash-based MAC) to ensure the received data hasn't been tampered. HMAC is a lightweight cryptographic hash function. However, all clients who are aware of the secret key can sign or verify the data with the hash [8].

## III. PREVALENT NETWORK VULNERABILITIES AND CYBER THREATS

In this section, the nine most prevalent attacks in SCADA IIoT systems and the associated risks are studied. It should be noted that due to their fundamentally different nature, the prevalent vulnerabilities and security priorities in ICSs are different from the ones in traditional IT systems. Since there has been an extensive discussion on these differences, and this matter is beyond the scope of this paper, we refer the readers to the references [9] and [10].

The CIA triad (Confidentiality, Integrity, and Availability) are the security traits that must be preserved in any system to keep the data safe. Further, the AAA security controls (Authentication, Authorization, and Accountability) are the security tools to protect the CIA traits in the system. In this section, we study the most prevalent attacks targeting each of these security elements. Accountability has been left out since it is generally an administrative aspect.

Our extensive study of relevant works (such as [9], [11], [12], [13], and [14]) reveals that these are the most common threats in the SCADA systems. However, unlike the existing works, in this paper, we

- provide a comprehensive set of prevalent attacks,
- define each attack separately regarding which security aspect they compromise,
- explain how they impact the IIoT performance,
- run a risk assessment based on the damage severity and the likelihood of happening in these specific systems, and
- study the effectiveness of machine learning (ML) based security solution to encounter each class of attack.

### *A. Prevalent Attacks*

The attacks are divided into five classes, based on which security aspects (integrity, availability, confidentiality, authentication, and authorization) are compromised. However, it is nearly impossible to define a solitary classification because the classes in which these attacks fall are not mutually exclusive. Often, compromising one aspect leads to compromising others as well.

#### *1) Integrity*

##### *1.a) Buffer Overflow*

In buffer overflow attacks, the intruder tries to write large data (more than the designated size) in the buffer, causing the extra bits to overflow and overwrite other buffers and alter their values. This attack is usually caused due to poor input type or size validation mechanisms and makes the system unreliable or even crash.

Buffer overflow attacks are highly prevalent in SCADA systems due to two main reasons. First, the majority of the operating systems in ICSs are written in programming languages such as C, which lacks type safety mechanisms. Further, SCADA devices operate continuously. The operating systems that have not been rebooted for years are more vulnerable due to accumulated memory fragmentation.

The buffer overflow problem in SCADA systems can affect both the supervisory control and field devices such as sensors. PLC's instructions to the output elements (e.g., turning on or off the water pumps) and sensed data (e.g., water level) could be manipulated through this attack [10].

##### *1.b) Code Injection*

In a code injection attack, the intruder tries to execute malicious commands or inject malicious data into the system. For instance, in a SQL injection attack, SQL queries are sent to control or compromise the database server. This attack exploits the system vulnerability due to the lack of user-supplied input data validation techniques.

This attack enables the intruder to access sensitive information such as usernames and passwords, and also alter the data (e.g., allowing access to an unauthorized user, deleting the data, etc.). A command injection attack can manipulate the control commands in the system and disrupt the normal operation.

Since the primary function of the SCADA systems is collecting and storing information, this attack may have a serious impact on the system. More specifically, if the system is controlled remotely through a web interface, this attack is able to compromise the data and the authentication procedures.

*1.c) Improper Input Validation*

This vulnerability is associated with the lack of proper mechanisms to validate the user's input. This is a more general type of vulnerability, which could lead to other types of risks. The attacker may be able to enter wrong values that can make the system unstable. Moreover, since these systems are not checked regularly due to their deterministic nature, this attack might stay undetected for a long time [12].

*2) Availability*

*2.a) Denial of Service (DoS)*

An intruder carries out a DoS attack to flood the targeted computer (e.g., PLC and HMI). This attack disrupts the availability of the SCADA system by sending a large number of random packets to the target node at a high rate to make the target unresponsive and may even crash the whole system.

A DoS attack against a SCADA system is generally carried out by an intruder connected to the network using SYN or HTTP flooding against a host. SYN attacks are constant fake synchronize requests, and HTTP attacks are either GET or POST requests to keep the web server of the target busy and not be able to respond to the normal traffic. If the links in the network are congested, monitoring and controlling the ICS will be highly difficult, if not impossible.

Therefore, the main goal of the DoS attack is to hurt the system's availability, so that legitimate users are not able to access the resources.

*3) Confidentiality*

*3.a) Reconnaissance*

In a reconnaissance attack, the intruder engages with the SCADA network to gather information about the system, such as the connected devices, security policies, IP addresses, host information, etc. After identifying the elements of the network, the attacker maps the network architecture to identify the vulnerabilities in the system. Eventually, the attacker may use this information to run exploits against susceptible devices to interrupt the system's functionality.

Intruders may start this attack using sniffers. They eavesdrop and inspect the ongoing network traffic to gain information about the network elements and their status. Stealth scan in SCADA network can occur on the link between any of the two nodes of the network; for instance, the link between the I/O network and the PLC or the link between the HMI computer and the PLC. This attack is considered passive since the attackers are silent and do not inject any traffic that would expose them. Although this attack may not be considered severe, the network information is exposed to an unauthorized person, and it is very difficult to detect.

*4) Authentication*

*4.a) Unauthenticated Access*

This vulnerability is due to poor authentication mechanisms in SCADA systems. Since these systems run continuously and autonomously, personnel may not change their usernames and passwords regularly. They may even use default usernames and passwords for ease of remembering [11]. Brute force methods or logging the user's keystrokes can be used to obtain this information. Furthermore, phishing attacks have been conducted widely to collect the credentials of ICS operators [12]. If the attacker somehow figures out these credentials, he can misuse his access and conduct other types of attack.

Since under this category, we solely consider "accessing" the data, in which usually root access is not granted. We have classified this attack as low impact. Otherwise, they will be categorized in more severe attack types such as directory traversal.

*4.b) Man-in-the-Middle*

In the man-in-the-middle attack, the intruder eavesdrops on the communication links and tries to compromise the messages between two nodes while the nodes think they are still talking to each other directly. For instance, the intruder may send malicious commands to the actuators pretending to be the PLC or send false responses from the sensors to the PLC. Further, the intruder may discard or manipulate messages. This type of attack will have a valid syntax code; hence, rule-based IDS will not be able to identify it from the message format [15]. This type of attack can be mostly prevented through encryption techniques.

*5) Authorization*

*5.a) Directory Traversal*

In this attack, the intruder tries to access the restricted directories or files that are supposed to be root access only. This vulnerability is due to poor filtering or validation mechanisms for user-supplied inputs. Poor directory listing control is another cause of these attacks. In this type of attack, the intruder will be able to download sensitive files and information from the system.

This attack often also results in compromising other vulnerabilities in SCADA systems such as confidentiality, since the attacker might access private files in the system. Proper input validation methods can prevent this type of attack.

*5.b) Backdoor*

In a backdoor attack, the intruder tries to find a way around the authentication process to enter the system. Through the backdoor access, the attacker can log into the system, reach all the data and files on the system, and execute commands. Backdoor installation on the victim system may be done by an insider. Once installed, it is very difficult to detect this type of attack, and it is considered highly dangerous since it grants the intruder full access to the system.

In the case of ICSs, some of the vendors and manufacturers have backdoor accounts into their products for remote support and updates [16]. This vulnerability puts the system in danger, and in case of a successful attack, all the SCADA data will be exposed to the intruder.

### B. Risk Assessment

As discussed in the previous subsection, different prevalent attacks have different severity and different rates of recurrence. We have built a risk assessment matrix for these vulnerabilities, which is shown in Figure 2. In this assessment, the impact and the likelihood of occurrence of the prevalent attacks are combined for integrated analysis. The matrix in this figure has been designed based on our study of the prevalent vulnerabilities and the severity of the risks associated with them. Vulnerabilities have been arranged in the matrix in the order of their likelihood and impact severity. It is important to mention that this order in the presented risk assessment matrix is based on our experience and judgment, and it might slightly differ from case to case or for different applications. However, it becomes a convenient tool for future studies.

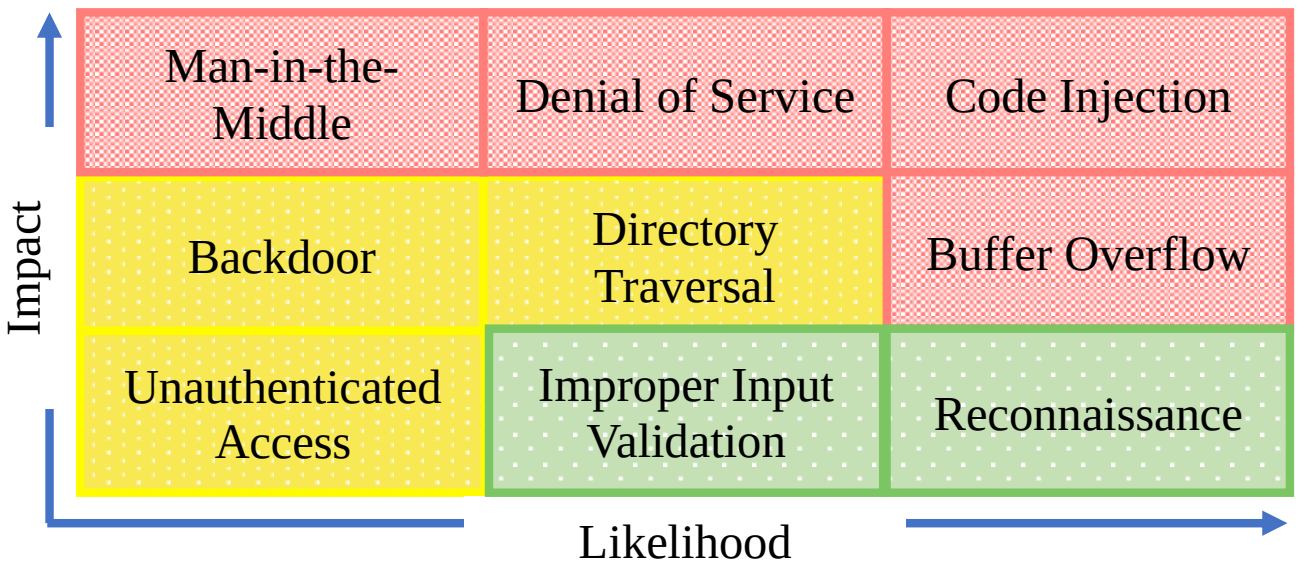

Figure 2. Risk Assessment Matrix of Prevalent Vulnerabilities in IIoT

Since we have picked the most common attacks, the likelihood is classified into three high levels of occurrence: occasional, likely, and certain. Similarly, the level of impact has been classified as mild, moderate, or critical. The overall risk ranking has been color coded. Threats that have severe negative impacts and are likely to occur frequently receive the highest rank, shown in the red color. Attacks with both low impact and low likelihood have the lowest rank, shown in the green color. And the yellow colored attacks fall between the two other classes. This risk assessment specific for the IIoT helps in identifying the threats that have the greatest overall risks and must be the top priority to address in these systems.

For instance, code injection is shown in red due to the catastrophic results of command manipulation in SCADA systems and their high probability of occurrence. DoS attacks that often occur would result in the termination of the system's operations. The reconnaissance attacks that harm the confidentiality may or may not lead to any negative consequences in the system's function.

### C. ML as a Versatile IIoT Security Tool

IDS has been widely used as an effective security mechanism to counter intrusions. Misuse-based IDSs such as rule-based, signature-based, flow-based, and traffic-based methods are just some examples of conventional IDSs. Since traditionally, most of the connections and traffic in the SCADA networks were pre-defined; these types of IDS were successful in detecting abnormal activities. For instance, when the intruder builds new connections to the victim or sends a different type of traffic, there will be unusual data flows in the network [17].

However, considering frequent upgrades in the networks, resulting in regular changes in the topology, the legacy IDSs do not perform properly. Also, to counter new types of attacks that appear every day, or in scenarios where the attack is planned intelligently (e.g., the man-in-the-middle attack), smart IDSs are required.

IDSs are, in general, helpful whenever the intruder affects the network data flow. This is true even for ML-based IDSs. If the intruder does not interact with any of the network elements, it is very difficult even to become aware of the intrusion. However, to launch attacks or compromise the network activities, the intruder has to disrupt the network somehow. The ability of ML algorithms in detecting small anomalies distinguishes them from any other type of IDSs.

ML algorithms can detect anomaly patterns that are difficult for humans to discover. To provide a secure network, the ML-based IDS can be designed with a moving target. This ability of ML models to learn and evolve is valuable because the attacks are constantly evolving, and new vulnerabilities are discovered every day. This is another reason why signature-based IDSs are becoming obsolete, and anomaly-based IDSs using ML are the new trend. We now discuss the suitability of ML-based IDS for each of the security elements.

*1) Integrity*

ML can be very helpful as a detection tool against data integrity threats. By training an ML-based IDS with legitimate traffic data, the IDS will learn the normal data that flow in the system. For instance, in the case of command injection, ML will detect the malicious queries that are out of the ordinary in the system. This specialized IDS is able to recognize the source that is compromising the integrity of the data to block him from the system to maintain the trustworthiness of the data. Hence, by learning the common behavior of the system, the ML-based IDS can be very useful against the attacks targeting this security element.

#### 2) *Availability*

ML can be very useful in detecting the DoS attacks. A proper ML algorithm can detect specific characteristics of the attacks targeting the availability, for instance, detecting the sources with unfamiliar or broadcast addresses, the ones that are showing abnormal behaviors, nodes that are sending an unreasonable amount of traffic, or when the normal operational traffic stops because the HMI or the PLC are flooded and unavailable.

Even though a simple network analyzer can detect the DoS attacks, it still requires a human operator to analyze the network logs. On the other hand, the ML-based IDS will not only provide proper automation but is also not prone to human error. Moreover, it has been shown to be effective in detecting this kind of out-of-ordinary behaviors.

#### 3) *Confidentiality*

In this type of attack, if the intruder merely eavesdrops on the network traffic (i.e., does not send any traffic nor build a connection with the devices in the network), it is very hard to detect using ML. As mentioned before, when the attacker's behavior does not change the network flow, it is very hard to detect the attack with any technique, including ML. However, as soon as the intruder engages with the network, a machine learning powered anomaly-based IDS will be able to recognize the abnormal behavior of the attacker trying to snoop or asking for unusual information from other nodes in the network. However, after engaging with the network, the malicious activities go beyond a simple eavesdrop attack and are classified under other attack categories.

#### 4) *Authentication*

As mentioned before, authentication is a security control technique. Attacks targeting this security element need to find a way around to bypass this step. To counter these threats, it will be more efficient to use prevention techniques rather than detection methods. For instance, encryption, strong passwords or key management techniques can be utilized to prevent unauthenticated access. Even though these techniques have their weaknesses, they improve the system’s robustness against unauthenticated access.

#### 5) *Authorization*

Activities that do not match with the normal traffic pattern even from verified users can be identified using ML techniques. Some examples include executing abnormal commands, manipulating the sensors and actuators, or sending random traffic on the network. If the intruder runs zero-day attacks or occasionally accesses the system, he might stay undetected for a while, but he will eventually be exposed by an ML-based IDS. However, the sensitivity of the learning technique must be high. The IDS learns the normal conditions of the system and will reveal abusive commands, unauthorized users, or intruders.

The IDS would raise the alarm each time it detects an abnormal behavior from a user in the network that must be verified by the operator. Raising the sensitivity of the utilized ML to detect these attackers will increase the number of false positives (normal traffic classified as attack traffic). Nevertheless, in security matters, it is better to be overcautious keeping the IIoT network safe. A false negative (undetected attack) could result in a higher cost than a false positive in the critical infrastructures.

## IV. EXISTING MACHINE LEARNING BASED IDSs

As we discussed in the previous section, since ICSs are different from regular IT systems, their communication type and even prevalent cyber vulnerabilities differ from a regular IT network. Consequently, it is important to consider these differences and design specific IDSs for SCADA systems.

In this section, we review available ML-based IDS approaches solving different security vulnerability issues of SCADA IIoT systems. Some of the presented research works focus on various security aspects. In this case, the first time that we mention their work, a detailed description is provided; following that in other subsections, just brief mentions are made.

### *A. Integrity*

Beaver et al. [18] employ six different types of ML algorithms, Naive Bayes, random forests (RF), OneR, J48, NNge (Non-Nested Generalized Exemplars), SVM (support vector machines). Their dataset consists of labeled remote terminal unit (RTU) telemetry data from a gas pipeline system in Mississippi State University’s Critical Infrastructure Protection Center. The attack traffic is generated from two types of code injection sets, command injection attacks, and data injection attacks. Seven different variants of data injection attacks were tried to change the pipeline pressure values, and four different variants of command injection attacks to manipulate the commands that control the gas pipeline.

Ullah and Mahmoud [19] suggest an IDS using a combination of J48 and Naive Bayes techniques. J48 is a type of DT (decision tree) technique. They have used the same dataset as the previously mentioned research work in [18]. The J48 classifier was first used as a supervised attribute filter. Then, the Naive Bayes classifier was used to develop the anomaly-based intrusion detection.

Alves et al. [20] employ the k-means technique, which is an unsupervised clustering algorithm. An open-source virtual PLC (OpenPLC platform) along with AES-256 encryption is used to simulate a SCADA system. They have conducted three different types of attacks against their system, code injection, DoS, and interception (eavesdrop).

He et al. [21] use CDBN (conditional deep belief network) to detect attacks in smart grids. They have simulated their system using IEEE 118-bus and 300-bus test systems. They also provide a comparison of their method with SVM and ANN (artificial neural network). They consider false data injection attack that is aimed at the integrity of data.

### B. *Availability*

Potluri et al. [22] design a hybrid IDS using SVM, and DBN (deep belief networks) for industrial networked control systems. They have used the NSL-KDD dataset, which is an old dataset and is not specific to ICSs but consists of DoS and integrity attacks.

As mentioned in Subsection A, paper [20] studies conducting DoS attacks as a part of their dataset to train their IDS.

### C. *Confidentiality*

Keliris et al. [23] use SVM in their simulated testbed in MATLAB controlling Tennessee Eastman (TE) chemical process. They have conducted reconnaissance attack as vulnerability discovery technique, and further, tried to study the effect of command injection attack on the controller to manipulate the reactor pressure.

As mentioned in Subsection A, paper [19] covers confidentiality problem through reconnaissance attack.

Alves et al. [20] (mentioned in Subsection A) also worked on eavesdropping attacks. Their unsupervised training methods were able to detect these attacks successfully.

### D. *Authentication*

The research work [20] was mentioned previously. The authors also declare that due to the utilization of encryption, their system is resistant to man-in-the-middle attacks.

Eigner et al. [24] employ KNN (k-nearest neighbors) on a custom-built conveyor belt system. They use the normal behavior of the system to design an anomaly-based attack detection approach. They study the performance of the system with different k's and different distance measurement metrics. They have focused on man-in-the-middle attacks as the intrusion scenario.

### E. *Authorization*

We could not find any work covering this area of intrusion detection for SCADA systems using ML techniques. Therefore, in this paper, we focus on unauthorized intrusion detection in our case study. More specifically, we have conducted backdoor attacks on our SCADA IIoT testbed. The details are provided in Section V.

### F. *No Specific Attack*

Yasakethu and Jiang [25] study the advantages and disadvantages of four different types of IDSs, rule-based, ANN, HMM (hidden Markov model), SVM, OCSVM (one-class SVM). However, they do not provide any implementation nor any practical analysis of these techniques.

Zhang et al. [26] have used SVM to detect anomalies in network traffic. They have generated their dataset using simulations of the IEEE 118 bus network. Six features of traffic (magnitude and phase of the current, magnitude, and phase of voltage, real power, and reactive power) were extracted from the data. In that paper, the normal condition was defined as when no equipment is disconnected from the system, and there is no fault in operation. The fault condition is defined as a short circuit occurrence somewhere in the system. No cyber-attack was conducted against the system.

Skripcak and Tanuska [27] have designed a multi-agent architecture for SCADA systems to monitor the plant processes using passive-aggressive on-line ML algorithms. The focus of the paper is to provide the theory behind forecasting based on the current situation.

Siddavatam et al. [17] employ DT and RF techniques. Their system prototype has been built in their lab. They have extracted several features from the traffic for training such as TTL (time to live), byte count for response type, word count for query type, packet type, and the reference number for query type. To generate abnormal behavior in the system, changes in operation were conducted through a control node, but no attacks were developed in the system.

Maglaras et al. [28] use OCSVM as their proposed anomaly-based IDS. They declare that OCSVM is a good choice because the dataset is imbalanced. The authors have used only two features of traffic (data rate and packet size) of an electric grid. The trained model did not include any malicious attack data, and the trained dataset was captured during normal operation of a SCADA system.

Mantere et al. [29] have focused just on feature selections in designing an anomaly-based detection. They have chosen features such as flow directions, individual packet sizes, protocol, average packet rates, average data byte rates as the most determining features. The traffic was captured from two different locations within an industrial site. However, no attack data were considered in this paper.

### G. *Summary*

Table I provides a summary of this section. In this table, the available ML-based anomaly detection approaches in SCADA are classified based on the type of their targeted vulnerabilities. This table provides a concise overview of where the most focus of the research works available in the literature is.

As shown in this table, we could not find any ML-based research work in authorization aspects of SCADA security. Hence, we have focused on this area in the next section.

Table I. Available ML-Based IDS for ICS Vulnerabilities

| | SVM or OCSVM | NB | DT or RF | DBN | ANN | KNN or K-means |
|---|---|---|---|---|---|---|
| Integrity | [18] [21] [22] [23] | [19] [18] | [19] [18] | [21] [22] | [21] | [20] |
| Availability | [22] | | | [22] | | [20] |
| Confidentiality | [23] | [19] | [19] | | | [20] |
| Authentication | | | | | | [20] [24] |
| Authorization | | | | | | |
| No Specific Attack | [25] [26] [28] | | [17] | | [25] | |

## V.OUR CASE STUDY

In this section, we describe our testbed and the evaluation results of our proposed ML-based IDS. The problem of intrusion detection in the IIoT systems is explored. To detect manipulated commands, system's transactions were logged and used to train the ML algorithms. We start this section by introducing our prior work and how this work is different, then the details on the testbed implementation, conducted cyber-attacks, and then the designed IDSs and their performance evaluations.

### *A. Our Prior Work*

In [30], we have presented the effect of imbalanced datasets. Even though ML has proven its capability in intrusion detection, there are cases that it falls short. Severely imbalanced training datasets where the number of attack data is significantly lower than normal data (e.g., less than 1%) is a real-world challenge that is quite common in IIoT security.

In [31] and [32], we have investigated IDS design using ML and ANN models for securing the confidentiality and availability (reconnaissance and DoS attacks) of the system.

However, in this work, we have improved our testbed by adding the following elements:

• Turbidity alarm and turbidity sensor have been embedded in the testbed to add analog input to the system. This helps us investigate overwriting analog registers (turbidity level of the water) in the PLC.

• Backdoor, command injection and SQL injection attacks are used to study the authentication and data integrity aspects of the security for these systems.

• New metrics have been used to evaluate the performance of the trained model. These metrics represent the quality of the performance with better granularity.

• Feature importance ranking study has been conducted to show which features are the most salient ones in distinguishing the attack traffic from the normal.

### *B. Our SCADA IDS Testbed Implementation*

Industrial companies almost never release their network data, because they are obligated to follow confidentiality laws and user privacy restrictions. Hence, real-world IIoT datasets are not available for security research in this area. In the IIoT security domain, the researchers usually have to use commercial or public datasets that are not specific to this domain. In our work, we developed a real-world testbed that resembles an actual industrial plant. We have conducted real cyber-attacks against the system to gather realistic datasets containing both normal and attack traffic analogous to real industrial network traffic.

We have picked an IIoT system that supervises the water level and turbidity quantity of the water storage tank, shown in Figure 3. This system is a part of the water treatment and distribution process in industrial reservoirs. This testbed includes components like historical logs, HMI, PLC. There are three sensors and four actuators in this testbed. Two water level sensors and an analog turbidity sensor compose the inputs. A three-light turbidity alarm, a valve, and two water are the actuators that receive the commands from the PLC. Also, there are control buttons (On, Off, Light Indicator) for manual control of the system. A detailed explanation of the testbed and its elements can be found in our previous work [30].

In short, the task of this testbed is to keep the water level between two pre-defined levels. At the same time, it measures the turbidity level of the water and illuminates one of the red, yellow or green lights of the turbidity alarm, based on the cloudiness level of the water. Modbus was utilized as the communication protocol in our testbed since it is one of the most popular IIoT protocols commonly used in the industrial control systems. The logic of the PLC is programmed using the Ladder language [33], [34].

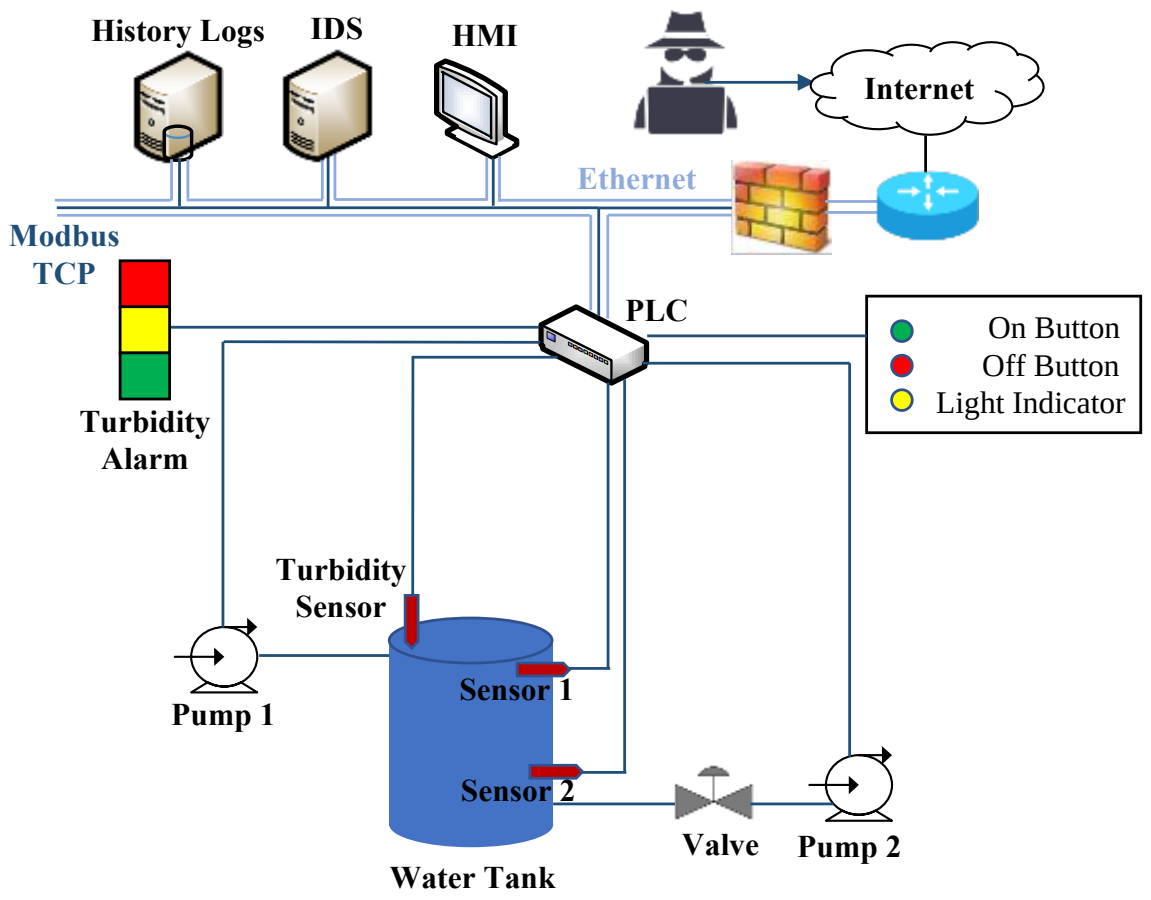

Figure 3. Scheme of Our Implemented Testbed

### *C. Our Attack Scenarios*

Since, to the best of our knowledge, no research paper has focused on machine learning-based IDS in SCADA systems for backdoor attacks, we conducted these attacks along with two other types of cyber-attacks. We have generated SQL injection and command injection attacks to have a larger variety of attack records in our dataset. These attacks were carried out using the Kali Linux Penetration Testing Distribution [35]. All the data generated during the attack phase as well as the normal traffic was gathered and recorded by Argus [36] and Wireshark [37] network tools.

An important point that should be mentioned here is that we have deliberately built our dataset to be imbalanced. The percentage of attack traffic in the dataset is less than 0.2%. This assumption makes the system as similar as possible to the real-world industrial control systems. The statistics of the dataset are shown in Table II, where the average data rate was 419 kbit/s, and the average packet size was measured as 76.75 bytes.

Table II. Our Built Dataset Statistical Information

| Type of the Traffic | Percentage |
|---|---|
| Normal Traffic | 99.81 |
| Total Attack Traffic | 0.19 |
| Backdoor Traffic | 0.085 |
| SQL Injection Traffic | 0.065 |
| Command Injection Traffic | 0.04 |

A brief explanation of how these three attacks disrupt the normal operation is provided next.

*1) Backdoor*

In this attack, our target is the HMI system, which gets infected with a backdoor virus. This virus works in the background and is hidden from the SCADA system's operator. The backdoor virus opens a port in HMI allowing a remote connection to be established with the attacker's PC. Thus, the attacker gains full access to the HMI computer, where the SCADA system is installed. Using the backdoor, the attacker can explore the HMI system and download any file, including the dataset with all the sensor and actuator values. In this attack scenario, we (as the white hat attacker) transferred about 1GB of files containing sensitive information from the HMI to the attacker's PC. We ran this attack several times to get the status of the system in different situations. Further, using this attack, we built new directories in the SCADA system and removed several files to disrupt the HMI operation.

*2) Command Injection*

In this attack, the target is the PLC. First, the attacker's PC connects to the network and is able to read all the PLC register values and logs them into a .txt file. After gaining access to the PLC register information, the attacker rewrites some of the PLC registers that are vital to the physical process. For example, while Pump 2 was supposed to draw water from the tank, we (as the white hat attacker) stopped it, started Pump 1, and the water flowed out from the tank. Another instance is when we turned on the wrong turbidity alarm light, in the way that, while the turbidity level was high, and the red light was supposed to be on, we (as the attacker) turned off the red light and turned on the green light instead.

*3) SQL Injection*

In this attack, both the HMI system and the PLC device are targeted. They both have web servers for setting up and accessing their configuration and information. In this attack, the attacker's PC sends database queries to submit untrusted data. We (as the white hat attacker) sent malicious SQL commands which were executed on the database. We ran this attack many times and logged all the network traffic.

### *D. Feature Selection*

An important step in training the ML models is selecting and extracting features from the traffic. Here, in designing our IDS, we chose the features that their values change during the attack phases compared to the normal operation phases. If a selected feature does not vary during the attacks, then even the best algorithm will not be able to detect an intrusion or an anomalous situation using that feature.

In our study, we reviewed the potential features and chose 23 features that are common in network flows and also change during the attack phases. Table III shows the chosen features along with their description.

Table III. Selected Traffic Features in Our Proposed IDS

| Features | Type | Descriptions |
|---|---|---|
| Mean flow (mean) | Float | The average duration of the active flows |
| Source Port (Sport) | Integer | Source port number |
| Destination Port (Dport) | Integer | Destination port number |
| Source Packets (Spkts) | Integer | Source/Destination packet count |
| Destination Packets (Dpkts) | Integer | Destination/Source packet count |
| Total Packets (Tpkts) | Integer | Total transaction packet count |
| Source Bytes (Sbytes) | Integer | Source/Destination bytes count |
| Destination Bytes (Dbytes) | Integer | Destination/Source bytes count |
| Total Bytes (TBytes) | Integer | Total transaction bytes count |
| Source Load (Sload) | Float | Source bits per second |
| Destination Load (Dload) | Float | Destination bits per second |
| Total Load (Tload) | Float | Total bits per second |
| Source Rate (Srate) | Float | Source packets per second |
| Destination Rate (Drate) | Float | Destination packets per second |
| Total Rate (Trate) | Float | Total packets per second |
| Source Loss (Sloss) | Float | Source packets retransmitted/dropped |
| Destination Loss (Dloss) | Float | Destination packets retransmitted/dropped |
| Total Loss (Tloss) | Float | Total packets retransmitted/dropped |
| Total Percent Loss (Ploss) | Float | Percent packets retransmitted/dropped |
| Source Jitter (ScrJitter) | Float | Source jitter in millisecond |
| Destination Jitter (DrcJitter) | Float | Destination jitter in millisecond |
| Source Interpacket (SIntPkt) | Float | Source interpacket arrival time in millisecond |
| Destination Interpacket (DIntPkt) | Float | Destination interpacket arrival time in millisecond |

How each feature varies depends on the type of attack. For instance, during the normal condition, where there is no attack, the SrcPkts and DstPkts features mostly show a periodic behavior. On the other hand, during attacks, these features show random behavior.

Further, we have studied the importance of the features. They are ranked based on how salient they are in helping the algorithm distinguish the normal traffic from the attack traffic. In this technique, the values of each feature are permuted randomly one at a time, creating new datasets. The ML model is trained on these datasets, and the increase in classification error is measured for each. If the increase is high, then the feature is important, and conversely, if it is low, the feature is considered as not important. For each feature, the "model reliance" or importance coefficient is defined as the ratio of the model's error value after permutation to the standard error value when none of the variables are permuted. For more detailed information, we refer readers to [38] and [39].

As we report later in this Section, random forest (RF) has shown the best classification performance, so we have picked this algorithm to calculate the importance. In Figure 4, the top five important features in our dataset along with their normalized (so the total of 23 feature importance values sum to 1) importance coefficient are shown. While these are the top five features, the threshold for the importance has shown that all the 23 features are required for training.

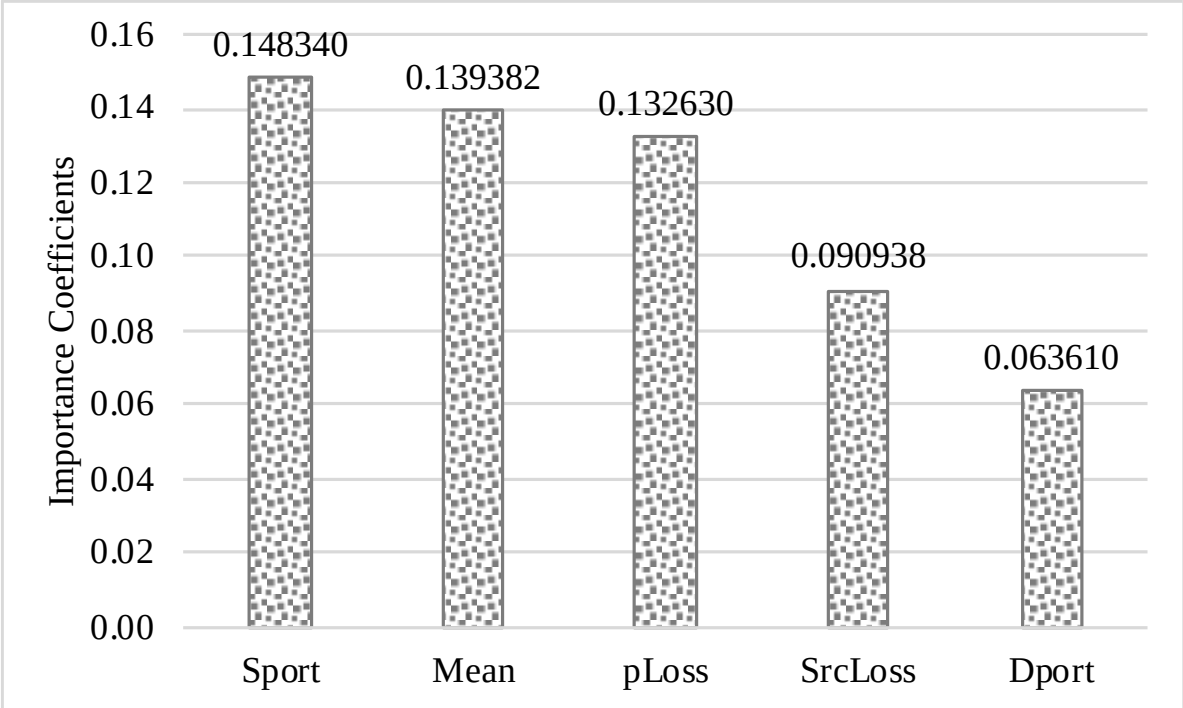

Figure 4. Top Five Important Features

### *E. Machine Learning Techniques*

In this case study, the ML-based IDS is designed just as a binary classification to decide whether a particular traffic sample is an attack or normal. The inputs to the IDS are the 23 chosen features, as mentioned in the previous section, and the output of the IDS is either 0 (normal traffic) or 1 (attack traffic). Also, for a total of 451,372 traffic samples, we use the ratio of 80% to 20% to divide the dataset into training and testing sets, respectively.

We have used and tested seven different techniques for the IDS; SVM, KNN, Naive Bayes (NB), RF, Decision Tree (DT), Logistic Regression (LR), and Artificial Neural Network (ANN). We have used the Keras library [40] to build the ANN, and for the other algorithms, the scikit-learn library [41] was utilized to develop the learning models for the IDS. The models are trained and tested over the data collected in the testbed, and the results of their performance are compared.

### *F. Performance Metrics*

Table IV. Confusion Matrix in IDS Context

| | | Predicted Class | |
|---|---|---|---|
| | | Classified as Normal | Classified as Attack |
| Actual Class | Normal Data | True Negative (TN) | False Positive (FP) |
| | Attack Data | False Negative (FN) | True Positive (TP) |

Traditionally, the performance of the ML algorithms is measured by metrics which are derived from the confusion matrix. Table IV shows the confusion matrix. The description of the matrix confusion parameters is as follows:

- True Negatives (TN): Represents the number of normal packets correctly classified as normal.
- True Positives (TP): Represents the number of abnormal packets (attacks) correctly classified as attacks.
- False Positive (FP): Represent the number of normal packets incorrectly classified as attacks.
- False Negative (FN): Represents the number of abnormal packets (attacks) incorrectly classified as normal packets. As an example, we picked the RF model to show its classification results in the form of confusion matrix in Table V.

Table V. Confusion Matrix of RF Classification Results

| | | Predicted Class | |
|---|---|---|---|
| | | Classified as Normal | Classified as Attack |
| Actual Class | Normal Data | 450561 | 30 |
| | Attack Data | 20 | 761 |

Based on the confusion matrix, the metrics used in this work to evaluate the performance of the ML algorithms are as follows:

- Accuracy: Shows the percentage of the correctly predicted samples considering the total number of predictions.

$$Accuracy = \frac{TP + TN}{TP + TN + FP + FN} \times 100 \quad (1)$$

- False Alarm Rate (FAR): Represents the percentage of the regular traffic misclassified as attacks.

$$FAR = \frac{FP}{FP + TN} \times 100 \quad (2)$$

- Undetected Rate (UR): The fraction of the anomaly traffic (attack) misclassified as normal.

$$UR = \frac{FN}{FN + TP} \times 100 \quad (3)$$

- Matthews Correlation Coefficient (MCC): Measures the quality of the classification. MCC shows the correlation agreement between the observed values and the predicted values.

$$MCC = \frac{TP \times TN - FP \times FN}{\sqrt[2]{(TP + FP) \times (TP + FN) \times (TN + FP) \times (TN + FN)}} \times 100 \quad (4)$$

- Sensitivity: Also known as the true positive rate. A sensitive algorithm helps rule out an attack situation with more confidence when the predicted data is labeled as "normal." While Sensitivity and the UR are complementary, each shows a different aspect of performance interpretation. If the focus is on minimizing FN, we would want to increase the Sensitivity of the model as much as possible (close to 100%), so that a smaller number of attacks stay undetected. Meanwhile, UR represents the fraction of these FN samples.

$$Sensitivity = \frac{TP}{TP + FN} \times 100 \quad (5)$$

Accuracy (Eq. 1) is the most frequently used metric for assessing the performance of binary classifiers. However, this metric is not sufficient for evaluation in scenarios with imbalanced classes (i.e., one class is dominant and has more training data compared to the other). In our case, which is an IDS scenario, the proportion of normal traffic to attack traffic is very high resembling a realistic dataset. This case is also valid where detecting rare anomalies is crucial like fraudulent bank transactions and identification of rare diseases. Therefore, in addition to the accuracy, we use other metrics evaluating the performance in a more meaningful way.

### *G. Results*

In this section, we present the numerical results of our algorithms detecting the attacks described in Subsection V.C. Figure 5 shows the accuracy results (Eq. 1). While RF shows the best performance and NB the worst, accuracy is not the best metric to evaluate the performance. As it was mentioned before, in scenarios like intrusion detection, the algorithms are biased toward estimating all the samples as normal. Even if an algorithm detects all the samples (even the attack ones) as normal, the accuracy will still be high, since the attack samples consist of a very small part of the dataset.

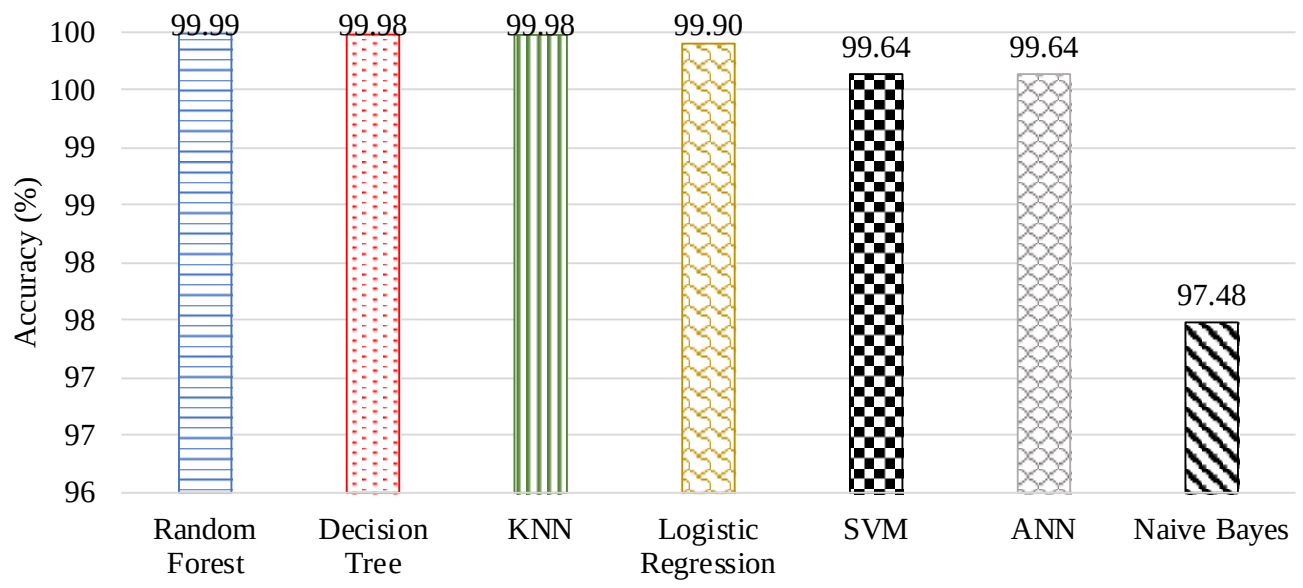

Figure 5. Accuracy

The false alarm rate (FAR), shown in Figure 6, represents the percentage of the normal traffic being misclassified as the attack traffic by the model (Eq. 2). Figure 6 shows good performance for all the models except NB. However, even this metric alone cannot truly represent the performance. Since the number of normal traffic is considerably higher than the attack data, and also the models are biased to label almost all the test data as normal (due to the imbalanced training dataset), the FAR value is expected to be low.

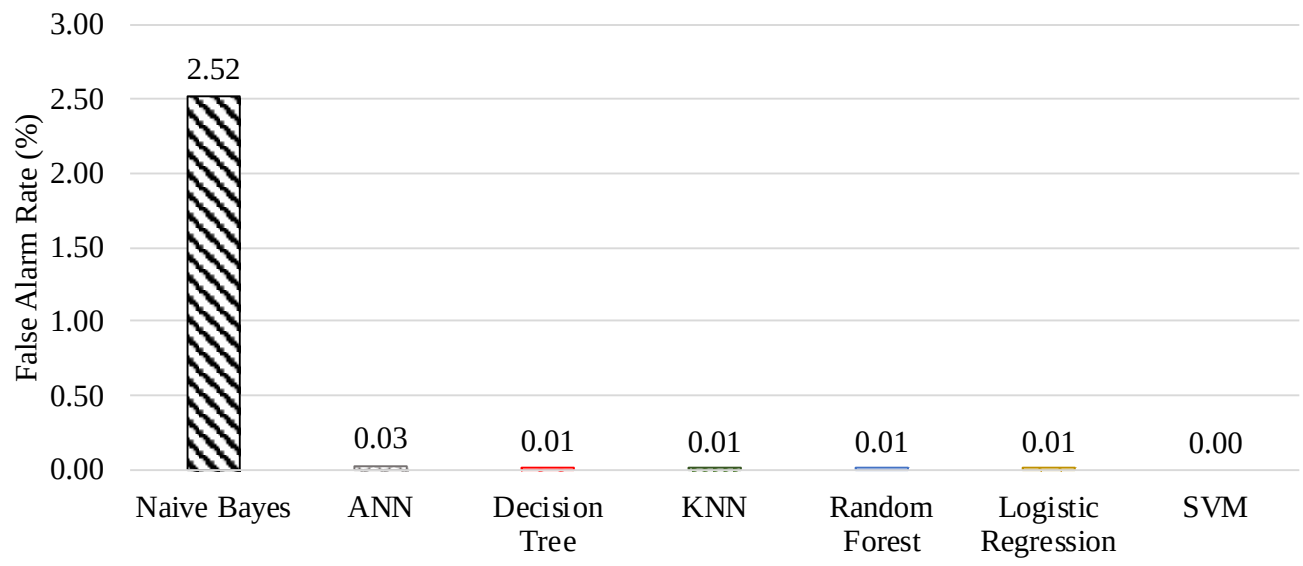

Figure 6. False Alarm Rate

Undetected rate (UR) metric can assess the performance better in spite of the imbalanced data. As shown in Figure 7, UR represents the percentage of the attack traffic that is misclassified as normal (the opposite of the FAR) (Eq. 3). Since this metric considers only the attack traffic, the fact of having an imbalanced dataset does not impact the evaluation. LR has the worst performance, even compared to a detector that would randomly assign true and false to each traffic packet, which would lead to 50% UR with an infinite number of packets. However, RF showed the best performance. This metric is more critical than FAR because it is related to the attacks not being detected by the system.

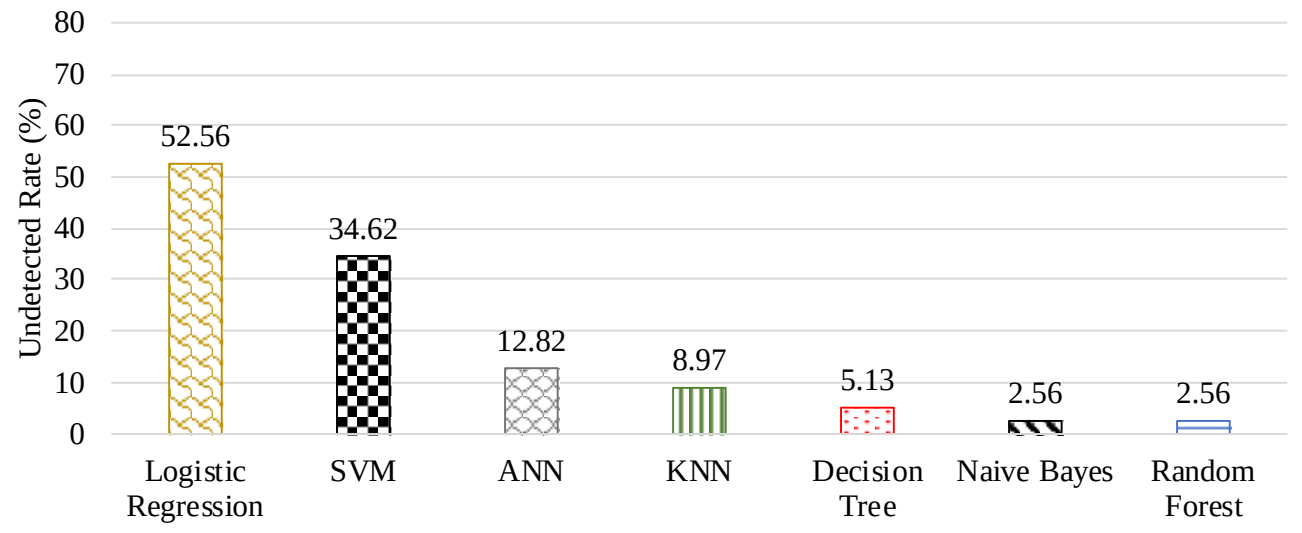

Figure 7. Undetected Rate

Figure 8 shows the ROC (receiver operating characteristic) curve. This curve basically plots the TP rate versus the FP rate for each model. As depicted, while RF shows the best performance, LR has the worst performance. The poor performance of LR for this metric is due to the low TP rate of the model in detecting abnormal traffic.

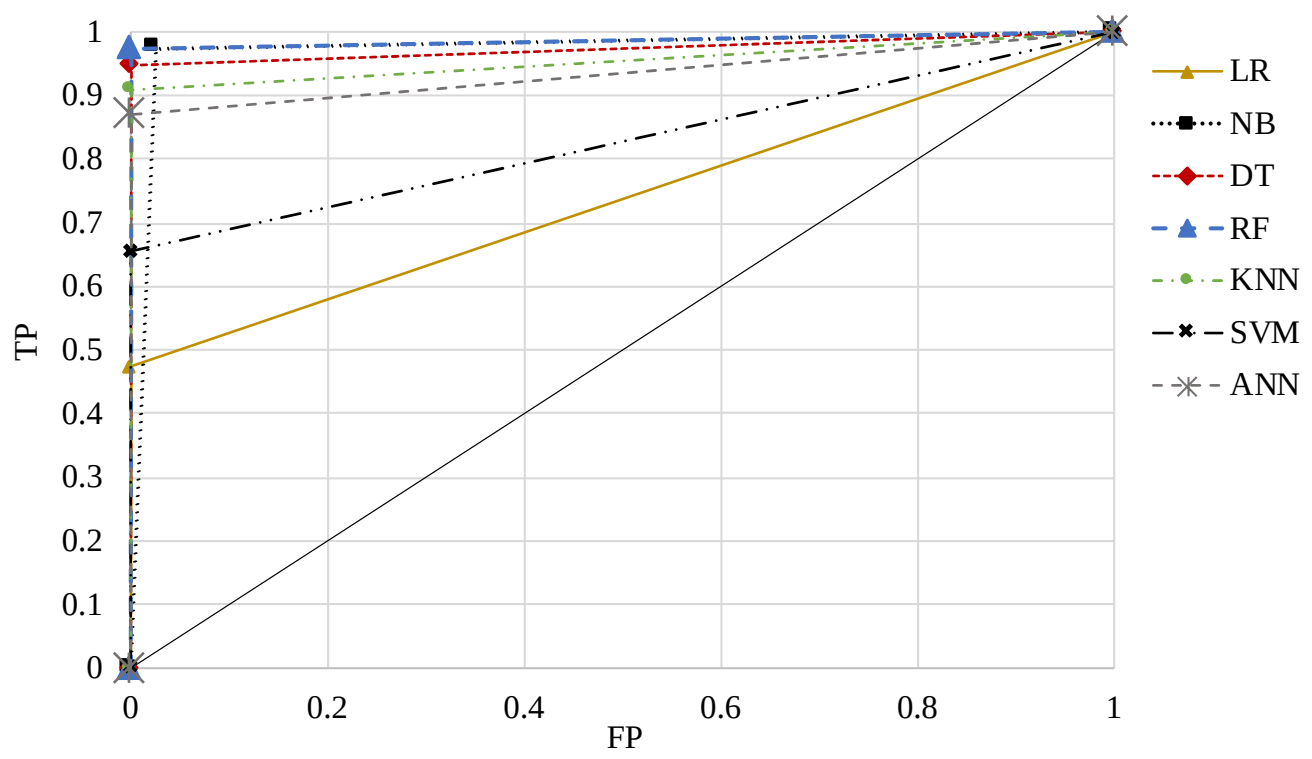

Figure 8. ROC Curve

MCC (Eq. 4), shown in Figure 9, is considered to be one of the best metrics for classification evaluation, and it is generally a better performance representative compared to the ROC curve and other metrics. As shown in this figure, RF has the best MCC value, while NB has the worst. MCC is considered as a fair metric when it comes to evaluating ML models that were trained with an imbalanced dataset. Since this metric represents the correlation agreement between the observed values and the predicted values, it is less affected by severe imbalanced ratios.

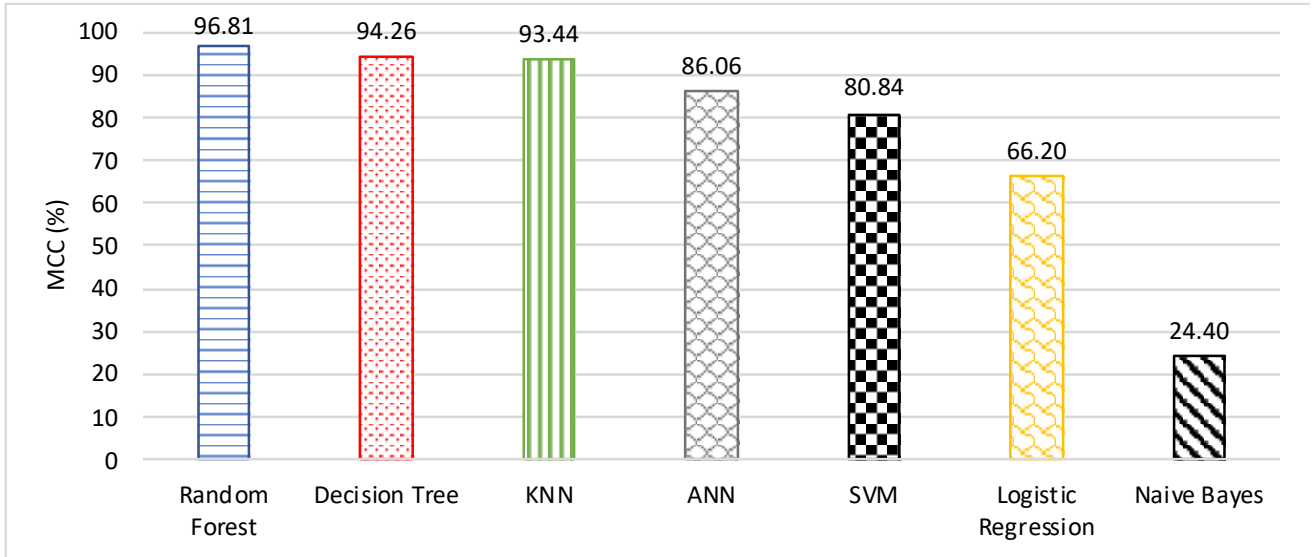

Figure 9. MCC

Finally, the sensitivity metric results (Eq. 5) are shown in Figure 10 to evaluate how sensitive each model is in reacting to an abnormal situation. As seen in the figure, RF and NB have the highest sensitivity, while LR shows the lowest.

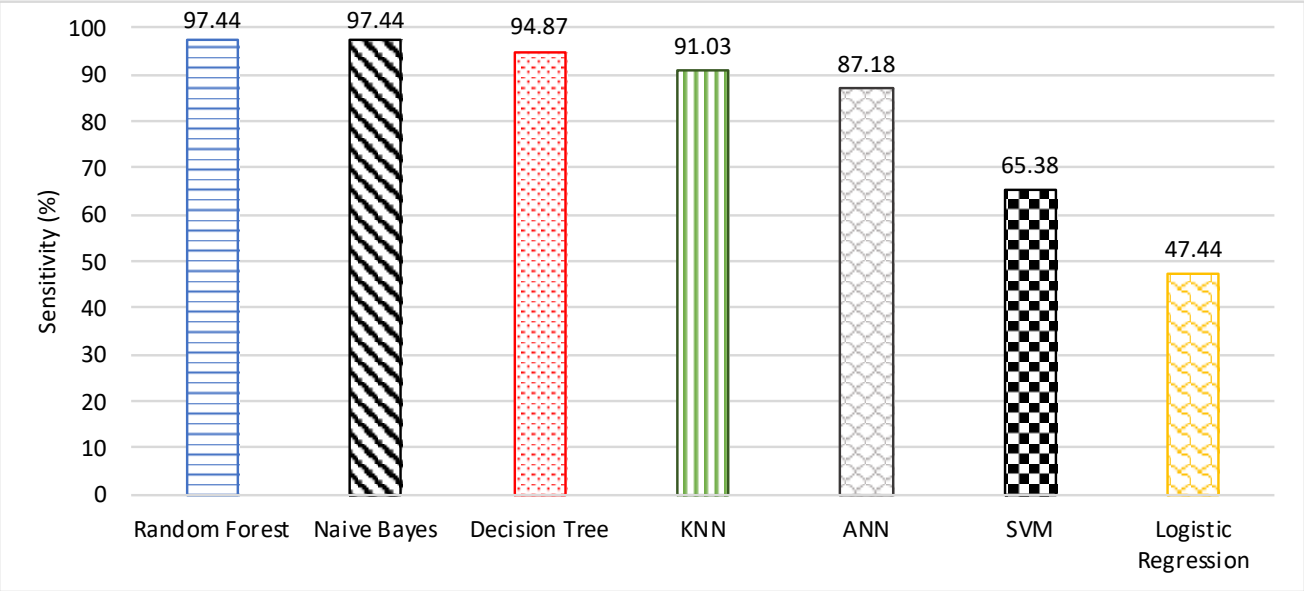

Figure 10. Sensitivity

## VI. CONCLUSIONS AND FUTURE DIRECTIONS

The cyber-security of the IIoT devices is critical. There is still a huge gap in providing adequate security for these systems, which is why it is crucial to focus on the industrial aspect of IoT technology. Machine learning solutions and big data analytics have been widely used to ensure a secure platform in the IT systems. However, due to the fundamental differences and dissimilar priorities of ICS and the traditional IT systems, their prevalent cyber-risks are different. Thus, special attention is required to provide security for IIoT. Through our discussions and experimental evaluation, we have demonstrated the effectiveness of machine learning for the security of these systems.

In this paper, we first studied the four most common protocols used in SCADA IIoT along with their security susceptibilities. Afterward, we carried out a risk assessment of the most important and prevalent vulnerabilities of the SCADA IIoT systems, and how ML-based solutions would be useful to combat them. Following that, a literature review on the existing anomaly detection approaches for SCADA systems using machine learning was provided to show where there is still a need for providing security. In the last section, we presented our case study and presented how machine learning is capable of filling the identified gap by handling new types of attacks such as backdoor, command injection and SQL injection. Feature importance ranking was also studied to highlight the most salient features in distinguishing the attack traffic from the normal traffic. The testbed built for this research work was designed to be as similar as possible to real-world IIoT scenarios. Special attention was also paid to evaluate the performance of the system using better representative metrics.

As our future direction, we plan to focus on utilizing a joint design of multiple algorithms to achieve better performance. The hybrid model should be able to provide more accurate results compared to any of the constituent models. False negatives, even a low number of them, mean malicious exertions against the system that stayed undetected and could lead to catastrophic results. Hence, reducing the number of false negatives is what we plan to concentrate on.

## VII. ACKNOWLEDGMENT

This publication was made possible by NPRP grant #NPRP 10-901-2-370 from the Qatar National Research Fund (a member of Qatar Foundation). The statements made herein are solely the responsibility of the authors.

Dr. Marcio A. Teixeira was supported by grant# 2017/01055-4 from São Paulo Research Foundation (FAPESP). FAPESP. He would also like to thank his primary institution - Instituto Federal de Educação, Ciência e Tecnologia de São Paulo (IFSP).

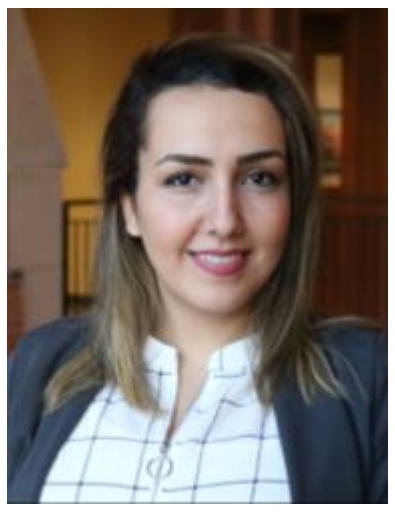

**Maede Zolanvari** is an IEEE student member. She received her B.S. and M.S. degree in Electrical and Computer Engineering, in 2012 and 2015 respectively. She's currently a Ph.D. candidate in Computer Science and Engineering at Washington University, St. Louis, MO, USA. During 2012 through 2015, her research was on performance improvement of communication networks. Since 2015, she has been working as a graduate research assistant at Washington University. Her current research focus is on utilizing machine learning and deep learning for network security of the Industrial Internet of Things. Her research interests include the Internet of Things, machine learning, cyber-security, secure computer networks, and wireless communications.

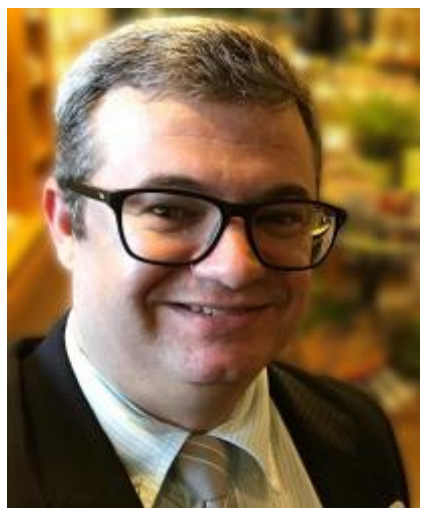

**Marcio A. Teixeira** is an IEEE senior member. He received his Ph.D. in Electrical Engineering at the Federal University of Uberlândia, Brazil in 2012, and received his MSc degree in Computer Science at the same university in 2004.
Currently, Dr. Marcio is a professor at the Federal Institute of Education, Science and Technology of São Paulo, Brazil. From 2017 to 2018, he worked as a postdoctoral researcher in Cybersecurity in the Department of Computer Science and Engineering at Washington University in St. Louis, MO, USA. His present research interest includes cybersecurity, network security, machine learning, deep learning, network performance analysis, wireless network protocols, and next-generation wireless communications.

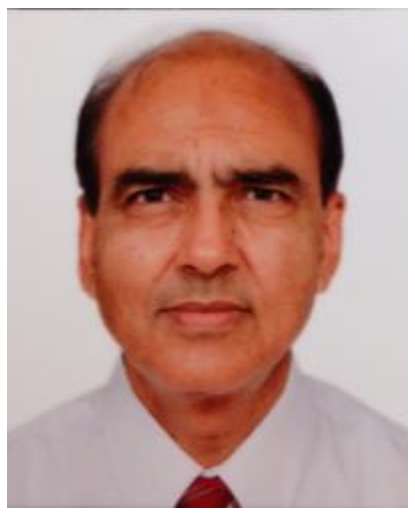

**Lav Gupta** is a senior member of IEEE. He received BS and MS degrees from the Indian Institute of Technology (IIT) in 1978 and 1980, respectively. He is currently a doctoral candidate in Computer Science and Engineering at Washington University in St. Louis, Missouri, USA. He has worked for about fifteen years in network planning, deployment, and regulation. He has also worked as a senior faculty of Computer Science and Access Network Planning in India and the UAE for a total of about fifteen years. He is the author of one book, twelve first author papers and has been a speaker at many international seminars. His current research areas are virtual network services, multi-cloud systems, fault and performance management in cloud-based, Cybersecurity and applied machine/deep learning. He is a recipient of best software award from Computer Society of India and best faculty award at Etisalat Academy, UAE.

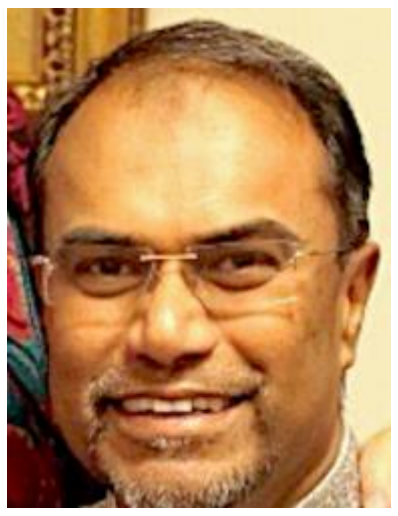

**Khaled M. Khan** is an associate professor in the department of Computer Science and Engineering and associated with the KINDI Centre for Computing Research at Qatar University. Prior to these, he served Western Sydney University (Australia) as a senior lecturer and was the Head of graduate programs. He received his Ph.D. in computing from Monash University, and BS and MS degrees both in computer science from the Norwegian University of Science and Technology. He is the Editor-in-Chief Emeritus of the International Journal of Secure Software Engineering.

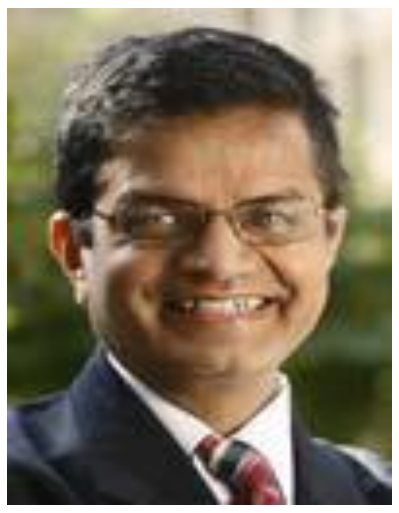

**Raj Jain** is a Fellow of IEEE, a Fellow of ACM, and a Fellow of AAAS. He received BS degree in Electrical Engineering from APS University in Rewa, India in 1972 and MS in Automation from IISc, Bangalore, India in 1974 and the Ph.D. degree in Applied Math/Computer Science from Harvard University in 1978. He is currently the Barbara J. and Jerome R. Cox, Jr., Professor of Computer Science and Engineering at Washington University in St. Louis. Previously, he was one of the Co-founders of Nayna Networks, Inc - a next generation telecommunications systems company in San Jose, CA. He was a Senior Consulting Engineer at Digital Equipment Corporation in Littleton, Mass and then a professor of Computer and Information Sciences at Ohio State University in Columbus, Ohio. Dr. Jain is the winner of the 2017 ACM SIGCOMM Life-Time Achievement Award. With 30,000+ citations on Google Scholar, he ranks among the Most Cited Authors in Computer Science.